\documentclass[amsmath,amssymb,prb,preprintnumbers,superscriptaddress,twocolumn]{revtex4}

\usepackage{ulem} 
\usepackage{epsfig}
\usepackage{color}

\usepackage{graphicx}
\usepackage{latexsym}
\usepackage{amsmath}
\usepackage{amssymb}
\usepackage{amsfonts}
\usepackage{bm}         

\newcommand{\gapprox}{{\scriptscriptstyle\stackrel{>}{\sim}}}

\newif\ifgraph

\graphtrue                   %

\begin{document}

\title{Local Tunneling Magnetoresistance probed\\ by Low-Temperature Scanning Laser Microscopy}

\author{Robert Werner}
\affiliation{Physikalisches Institut and Center for Collective Quantum Phenomena in LISA$^+$, Universit\"{a}t T\"{u}bingen, 72076 T\"{u}bingen, Germany}

\author{Mathias Weiler}
\affiliation{Walther-Mei{\ss}ner-Institut, Bayerische Akademie der Wissenschaften, 85748 Garching, Germany}
\author{Aleksandr Yu.~Petrov}
\affiliation{CNR-IOM TASC National Laboratory,S.S. 14 Km 163.5 in AREA Science Park, 34012 Basovizza, Trieste, Italy}
\author{Bruce A.~Davidson}
\affiliation{CNR-IOM TASC National Laboratory,S.S. 14 Km 163.5 in AREA Science Park, 34012 Basovizza, Trieste, Italy}
\author{Rudolf Gross}
\affiliation{Walther-Mei{\ss}ner-Institut, Bayerische Akademie der Wissenschaften, 85748 Garching, Germany}
\author{Reinhold Kleiner}
\affiliation{Physikalisches Institut and Center for Collective Quantum Phenomena in LISA$^+$, Universit\"{a}t T\"{u}bingen, 72076 T\"{u}bingen, Germany}
\author{Sebastian T.~B.~Goennenwein}
\affiliation{Walther-Mei{\ss}ner-Institut, Bayerische Akademie der Wissenschaften, 85748 Garching, Germany}%
\author{Dieter Koelle}
\email[Electronic address: ]{koelle@uni-tuebingen.de}
\affiliation{Physikalisches Institut and Center for Collective Quantum Phenomena in LISA$^+$, Universit\"{a}t T\"{u}bingen, 72076 T\"{u}bingen, Germany}

\date{\today}
\begin{abstract} 
Tunneling magnetoresistance (TMR) in a vertical manganite junction was investigated by low-temperature scanning laser microscopy (LTSLM) allowing to determine the local relative magnetization $\bm M$ orientation of the two electrodes as a function of magnitude and orientation of the external magnetic field $\bm H$.
Sweeping the field amplitude at fixed orientation revealed magnetic domain nucleation and propagation in the junction electrodes.
For the high-resistance state an almost single-domain antiparallel magnetization configuration was achieved, while in the low-resistance state the junction remained in a multidomain state.
Calculated resistance $R_\mathrm{calc}(\bm H)$ based on the local $\bm M$ configuration obtained by LTSLM is in quantitative agreement with $R(\bm H)$ measured by magnetotransport.
\end{abstract} 
\maketitle


Tunneling magnetoresistance (TMR) is an important effect for spintronics\cite{Bader10} which has been vigorously investigated both theoretically\cite{Slonczewski89,MacLaren97} and experimentally\cite{Julliere75, Moodera95}, predominantly to develop new devices such as magnetic random access memories (MRAMs)\cite{Scheuerlein00}  based on magnetic tunnel junctions (MTJs).
In an MTJ, two ferromagnetic electrodes are separated by a thin insulating tunnel barrier.
According to the Julli\`{e}re model\cite{Julliere75}, the maximum TMR ratio is $\mathrm{TMR_J} \equiv (R_\mathrm{ap}-R_\mathrm{p}) / R_\mathrm{p} $, where $R_{\mathrm{ap}}$ and $R_\mathrm{p}$ is the resistance for antiparallel and parallel orientation of the magnetizations $\bm M$ of the two electrodes, respectively.
While integral TMR properties of MTJs have been investigated in detail\cite{Moodera99, Bowen07}, not much is known about the impact of their magnetic microstructure on the TMR properties.
However, both in view of applications and from a fundamental point of view, it is of high interest to identify the spatial dependence of the TMR on the magnetic properties of the electrodes.
Nucleation and growth of magnetic domains in ferromagnets has been the focus of many efforts, using techniques such as magneto-optical Kerr, magnetic force, spin-polarized scanning tunneling, spin-polarized scanning electron and photoemission microscopy.\cite{Freeman01}
Low-temperature scanning laser microscopy (LTSLM) has been used to visualize locally different resistive states in a quasi-1-dimensional La$_{0.67}$Ca$_{0.33}$MnO$_3$ thin film grain boundary junction, for which it has been shown that the obtained LTSLM signal is directly proportional to the local TMR ratio.\cite{Wagenknecht06}
These results suggested that LTSLM could also be useful to investigate TMR in vertical MTJs under typical bias conditions and in a wide range of temperatures and magnetic fields.

In this letter, we report on LTSLM imaging of resistive states in a planar La$_{0.65}$Sr$_{0.35}$MnO$_3$/SrTiO$_3$/La$_{0.65}$Sr$_{0.35}$MnO$_3$ (LSMO/STO/LSMO) heterostructure MTJ upon variation of direction and amplitude of the external in-plane magnetic field $\bm H$.
The TMR of this device was investigated in Ref.~[\onlinecite{Werner11}].
LTSLM probes changes in the tunneling conductivity induced by local thermal perturbation, which allows to infer the spatial distribution of the relative magnetization orientation of the two electrodes.
By varying $\bm H$, imaging of magnetic domain nucleation and propagation during the field-driven transitions between low- and high-resistive states is possible.

\begin{figure}[tbp]
\includegraphics[width=0.48\textwidth]{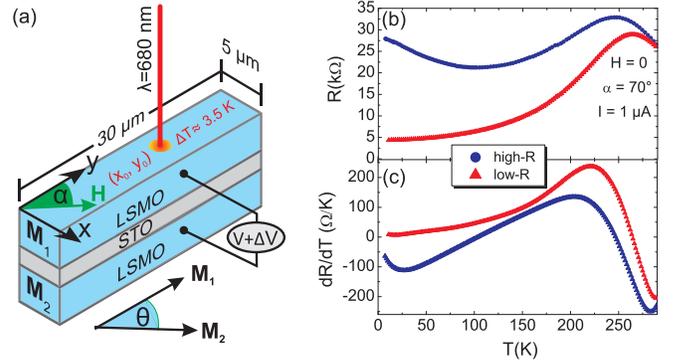}
\caption{(Color online) (a) MTJ geometry with angles $\alpha$ (between $y$ direction and in-plane field $\bm H$) and $\Theta$ (between magnetization $\bm M_1$ and $\bm M_2$).
(b) $R(T)$ for high-$R$ and low-$R$ states.
%
(c) ${\rm d}R/{\rm d}T(T)$ for same states as in (b).}
\label{fig:setup}
\end{figure}

\begin{figure*}[tbp]
\includegraphics[width=0.96\textwidth]{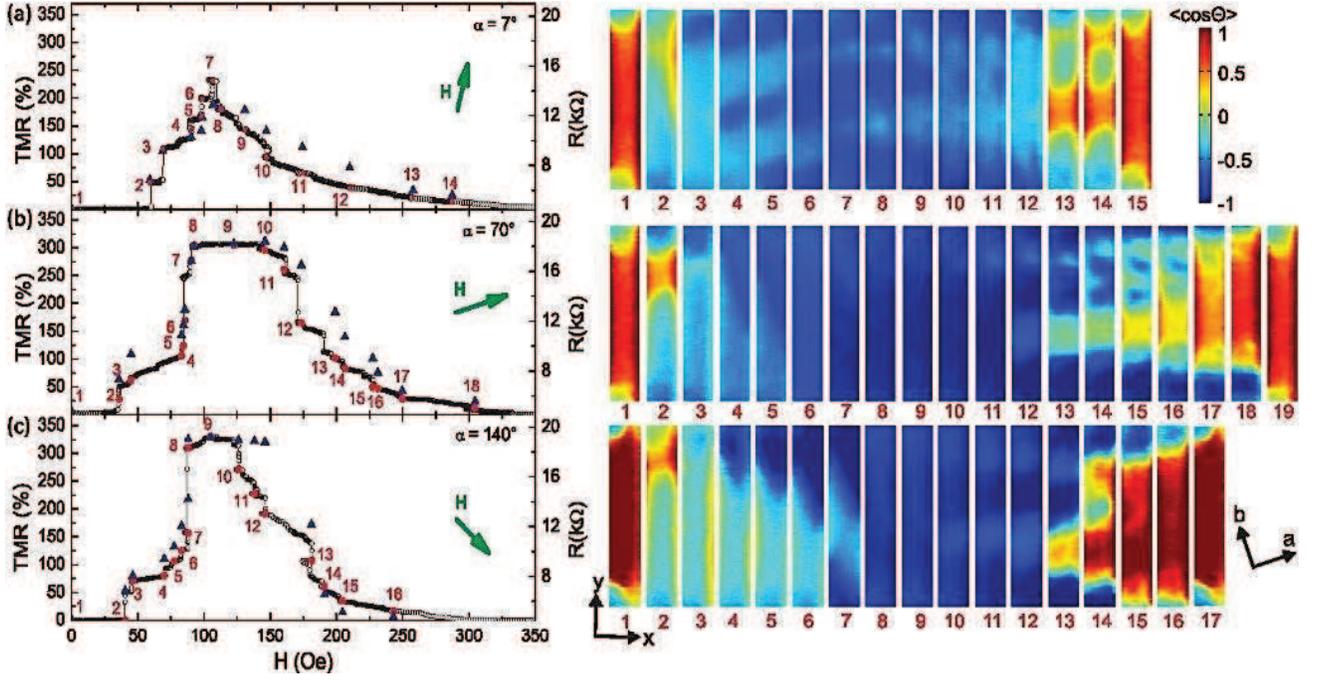}
\caption{(Color online) MTJ characteristics at $T$=30\,K ($I$=4\,$\mu$A) for different field orientation $\alpha$.
Left: $R$ and TMR vs $H$ (dots); triangles are  $R_\mathrm{calc}$  from corresponding LTSLM images.
Right: Spatial dependence of $\langle\cos\Theta\rangle$, as calculated from LTSLM voltage images for the bias points shown left and for $H=1\,$kOe [images 15 in (a), 19 in (b) and 17 in (c)].}
\label{fig:TTRLM}
\end{figure*}

The heterostructure samples were grown \textit{in situ} by molecular beam epitaxy on (001)-oriented STO substrates with 19\,nm
thick top and bottom LSMO electrodes, separated by
2.3\,nm thick STO.
A
38\,nm thick antiferromagnetic La$_{0.35}$Sr$_{0.65}$MnO$_3$ layer was grown underneath the junction trilayer to pin the bottom electrode via exchange bias.
MTJs  with area $A_J=5\times30\,\mu\rm{m}^2$ were patterned by optical lithography; for details on sample characterization and fabrication see Ref.~[\onlinecite{Werner11}].
Figure \ref{fig:setup}(a) shows a schematic view of the MTJ.
The relative difference in the orientation of the magnetization $\bm M_1$ and $\bm M_2$ of the top and bottom electrode, respectively, is given by the angle $\Theta$.
The angle $\alpha$ indicates the $\bm H$ direction with respect to the $y$-axis (long axis of the junction).

The sample was mounted on the cold finger of an optical $^{4}$He flow cryostat.
In-plane fields $|H|\le1\,$kOe were applied by a rotatable Helmholtz coil and the junction resistance $R$ was measured with constant bias current $I$ in  two-point configuration.
Figure \ref{fig:setup}(b) shows $R$ vs.~temperature $T$ warm-up curves at $H=0$, recorded after the junction was field-cooled at $\alpha=70^\circ$ to 5\,K and  either a low-$R$ state (with resistance $R_\mathrm{min}$) or high-$R$ state (with resistance $R_\mathrm{max}$) was prepared.

In LTSLM an amplitude modulated laser beam ($f$$\approx$10\,kHz, $\lambda$=680\,nm, $P$=5\,mW, spot diameter $1.5\,\mu$m~\cite{Wagenknecht06,Wang09}) is scanned across the sample.
Local heating in vicinity of the beam spot position $(x_0,y_0)$ on the sample surface induces a change in tunneling conductivity (conductance per area) $g$ and in turn a change $\Delta G$ in the global conductance $G$=1/$R$, which is detected as a change $\Delta V$ of the voltage $V$ across the sample.
$R(T)$ differs strongly for parallel and antiparallel orientation of the magnetizations, cf.~Fig.~\ref{fig:setup}(c), allowing to image local variations of $\Theta$; for a detailed analysis, see Ref.~[\onlinecite{LTSLM-signal}].
We assume that (i) local variations of $g$ are due to variations in $\Theta$, i.e.~arise from the formation of multidomain states in the electrodes and (ii) that the relation \cite{Jaffres01}
\begin{equation}
g(x,y,T)= g_p(T) \cos^2\{\frac{\Theta(x,y)}{2}\} + g_{ap}(T)\sin^2\{\frac{\Theta(x,y)}{2}\}
\label{Eq:1}
\end{equation}
holds, with $T$-dependent $g_p$=$g(\Theta$=$0)$ and $g_{ap}$=$g(\Theta$=180$^\circ)$.
Then from LTSLM images $\Delta V(x_0,y_0)$, we extract (cf.~Eq.(14) in [\onlinecite{LTSLM-signal}]) the convolution $\langle\cos\Theta(x_0,y_0)\rangle$$\approx$$a_1\Delta V(x_0,y_0)/R^2$+$a_2$
of $\cos\Theta (x,y)$ with the beam-induced $T$ profile\cite{Gross94}, centered at $(x_0,y_0)$ and assumed to be Gaussian with amplitude $\Delta T$ and variance $\sigma$.
Data analysis \cite{LTSLM-signal} yields
$\Delta T$=$3.5\,$K, $\sigma$=$0.8\,\mu$m, $a_1$$\equiv$$\frac{2A_J}{I(G_{ap}'-G_{p}')\Delta T 2\pi\sigma^2}$=$4.0\,\frac{(\mathrm{k}\Omega)^2}{\mu\mathrm{V}}$
and
$a_2$$\equiv$$\frac{G_{p}'+G_{ap}'}{G_{ap}'-G_{p}'}$=0.6; here, $G_p'$ and $G_{ap}'$ are the $T$-derivative of the MTJ conductance for homogeneous parallel and antiparallel  magnetization orientation, respectively.

For $T$=30\,K, Fig.~\ref{fig:TTRLM} shows $R(H)$ and the corresponding $\mathrm{TMR}(H)\equiv \left[R(H)-R_\mathrm{min}\right]/R_\mathrm{min}$ (left panel) together with images
$\langle\cos\Theta(x_0,y_0)\rangle$ for three different field orientations [increasing $\alpha$ from (a) to (c)].
Data were obtained for increasing field from $H$=0, after the junctions were prepared in the low-$R$ state upon applying $H$=$-$1\,kOe.
The overall shape of $R(H)$ depends strongly on $\alpha$.
However in all cases, upon increasing $H$, we observe a subsequent step-like increase in $R$ up to the maximum resistance $R_\mathrm{max}$, which depends on $\alpha$, and a further decrease in several steps down to $R_\mathrm{min}$ at $H \gapprox  300\,$Oe, which is roughly the same for all values of $\alpha$.
For all values of $\alpha$, at $H$=0 ($R$=$R_\mathrm{min}$) the LTSLM images reveal $\langle\cos\Theta\rangle$$\approx$1 except for the upper and lower edges where $\langle\cos\Theta\rangle$ approaches zero, indicative of domains with $\Theta$=90$^\circ$.
Consequently, the measured $R_\mathrm{min}$$\approx$4.5\,k$\Omega$ is about 2\,\% higher than our estimated value for $R_p$ (i.e.~for a homogeneous parallel orientation of the magnetizations)\cite{LTSLM-signal}.
For increasing $R(H)$ we find an inhomogeneous distribution of $\langle\cos\Theta\rangle$ and domains with $\Theta$$\sim$90$^\circ-110^\circ$ appear for all $\alpha$.
The maximum TMR value is reached for $\alpha$=140$^\circ$ and $H$=100 Oe.
The corresponding LTSLM image, Fig.~\ref{fig:TTRLM}(c) 9, reveals that the junction is almost completely in the antiparallel state here.
A similar value for $R_\mathrm{max}$ and a nearly homogeneous antiparallel configuration of $M_1$ and $M_2$ is found as well for $\alpha=70^\circ$.
By contrast, for $\alpha$=7$^\circ $, an inhomogeneous distribution of $\langle\cos\Theta\rangle$ can be seen even at maximum resistance and consequently the TMR is strongly reduced compared to $\alpha$=140$^\circ$  and $\alpha$=70$^\circ$.

We note that the observed magnetic texture is qualitatively different for, e.g., images 5 and 13 in Fig.~\ref{fig:TTRLM}(c), although the integral $R(77\,\mathrm{Oe})$=$R(180\,\mathrm{Oe})$.
This demonstrates that an integral determination of the TMR is not sufficient to extract information on the magnetic microstructure.
Furthermore, in the images of Fig.~\ref{fig:TTRLM}(b) and (c), domain walls are  predominantly oriented along the $a$-axis [cf.~Fig.~\ref{fig:TTRLM}(a)4-6 and Fig.~\ref{fig:TTRLM}(c)13-17] or $b$-axis [cf.~Fig.~\ref{fig:TTRLM}(b)4-5 and (c)4-7] of the LSMO electrodes.
From previous measurements of exchange-biased LSMO bilayers, the effect of exchange bias is primarily an increase of the coercive field of the pinned electrode (here, the MTJ bottom electrode) together with a small shift in the hysteresis loop.
This coercivity contrast is clearly visible in the $R(H)$ scans of Fig.~\ref{fig:TTRLM}(b) and (c), which show a large field window with nearly fully antiparallel magnetizations and a plateau in maximum TMR.
Therefore, the non-collinear magnetization configuration seen at small $H < 100\,$Oe results from domain switching in the top (unpinned) electrode, while the magnetic structure at large $H > 130\,$Oe results from domain switching in the bottom electrode.
This suggests that domains tend to form along the crystalline $b$-axis in the upper electrode [Fig.~\ref{fig:TTRLM}(c)4-7 and (b)4-6] and along the $a$-axis in the bottom electrode [Fig.~\ref{fig:TTRLM}(c)10-16 and (b)12-18].
Using LTSLM we can conclude that the switching process occurs via evolution of multidomain states, which form due to competition between crystalline anisotropy and shape anisotropy that can also be influenced by exchange bias.
Furthermore, even though the LTSLM signal reflects the local relative angle between the two magnetizations, under conditions of sufficient coercivity contrast, the LTSLM technique can also yield information about domain evolution in the individual electrodes.

For each image shown in the right panel of Fig.~\ref{fig:TTRLM}, using Eq.(\ref{Eq:1}) and $\langle\cos\Theta\rangle \approx \cos\Theta$ we can calculate the conductivity $g$ and, by integrating over the junction area, the conductance $G$ and junction resistance $R$, which is shown as $R_{calc}$ by triangles in Figs.~\ref{fig:TTRLM} (a)-(c).
The measured $R(H)$ and $\mathrm{TMR}(H)$ is reproduced quite well for values of $\alpha$ and $H$ below the TMR maximum.
For fields above the TMR maximum there are deviations from the measured curves; the origin of this we could not clarify yet.\cite{Note}

In conclusion, we have analyzed by LTSLM the local TMR of a LSMO/STO/LSMO MTJ.
In a quantitative analysis, we have calculated the local relative magnetization orientation of the two electrodes for different values of amplitude and direction of applied field.
LTSLM images visualized magnetic domain nucleation and propagation during magnetic field sweeps.
The domain walls are predominantly oriented along the crystalline $a$- and $b$-axes of LSMO.
The LTSLM images also allowed to calculate the global TMR, yielding quantitative agreement with the integral $R(H)$ measurements.
The results show that LTSLM can be used to link the magnetic microstructure to the integral magnetotransport properties and thus provides a valuable tool for further investigations of MTJs.

R.~Werner gratefully acknowledges support by the Cusanuswerk, Bisch\"{o}fliche Studienf\"{o}rderung.
This work was funded by the DFG (Projects KO 1303/8-1 andGO 944/3-1) and the German Excellence Initiative via the Nanosystems Initiative Munich (NIM).
B.~A.~D. and A.~Yu.~P. acknowledge support by the FVG Regional project SPINOX funded by Legge Regionale 26/2005 and Decreto 2007/LAVFOR/1461.


\begin{references}
\bibitem{Note}
\bibinfo{note}{Each image was taken in the following sequence: After preparing
  the MTJ at a certain value of $H$ (and measuring $R(H)$), the field was
  reduced to $H=0$, the image was taken, and the field was swept back to the
  initial value of $H$. During this sweep sequence, the resistance did slightly
  change (by typically $\pm 5-10\,\%$) but always returned back to the initial
  value. This procedure could partially explain the deviations between $R$ and
  $R_\mathrm{calc}$ for $\alpha=7^\circ$ for field values above the resistance
  maximum [cf.~Fig.~2(a)], as in this particular case the resistance upon
  sweeping back to $H=0$ went up by up to 20\,\%. However, this effect was much
  weaker in all other cases.}
\end{references}
\end{document}